\documentclass[twocolumn,notitlepage,superscriptaddress,showkeys,showpacs]{revtex4-1}
\usepackage{graphicx}
\usepackage{dcolumn}
\usepackage{booktabs,bm,color,braket}
\usepackage{color}
\usepackage[dvipsnames]{xcolor}
\usepackage[hidelinks,colorlinks,linkcolor=blue,
citecolor=blue,urlcolor=blue]{hyperref}
\begin{document}

\title{Giant room temperature anomalous Hall effect and magnetically tuned topology
in the ferromagnetic Weyl semimetal Co$_2$MnAl}
\author{Peigang Li}
\thanks{These two authors contributed equally}
\affiliation{Department of Physics and Engineering Physics, Tulane University, New Orleans, LA 70118, USA}
\author{Jahyun Koo}
\thanks{These two authors contributed equally}
\affiliation{Department of Condensed Matter Physics, Weizmann Institute of Science, Rehovot 7610001, Israel}
\author{Wei Ning}\email{wvn5038@psu.edu}
\affiliation{Department of Physics, Pennsylvania State University, University Park, Pennsylvania 16802, USA}
\author{Jinguo Li}
\affiliation{Superalloys Division, Institute of Metal Reseach, Chinese Academy of Sciences, Shenyang  110016, China
}
\author{Leixin Miao}
\affiliation{Department of Materials Science and Engineering, Pennsylvania State University, University Park, Pennsylvania 16802, USA}
\author{Lujin Min}
\affiliation{Department of Physics, Pennsylvania State University, University Park, Pennsylvania 16802, USA}
\affiliation{Department of Materials Science and Engineering, Pennsylvania State University, University Park, Pennsylvania 16802, USA}
\author{Yanglin Zhu}
\affiliation{Department of Physics, Pennsylvania State University, University Park, Pennsylvania 16802, USA}
\affiliation{Department of Physics and Engineering Physics, Tulane University, New Orleans, LA 70118, USA}
\author{Yu Wang}
\affiliation{Department of Physics, Pennsylvania State University, University Park, Pennsylvania 16802, USA}
\affiliation{Department of Physics and Engineering Physics, Tulane University, New Orleans, LA 70118, USA}
\author{Nasim Alem}
\affiliation{Department of Materials Science and Engineering, Pennsylvania State University, University Park, Pennsylvania 16802, USA}
\author{Chao-Xing Liu}
\affiliation{Department of Physics, Pennsylvania State University, University Park, Pennsylvania 16802, USA}
\author{Zhiqiang Mao}\email{zim1@psu.edu}
\affiliation{Department of Physics, Pennsylvania State University, University Park, Pennsylvania 16802, USA}
\affiliation{Department of Physics and Engineering Physics, Tulane University, New Orleans, LA 70118, USA}
\author{Binghai Yan}\email{binghai.yan@weizmann.ac.il}
\affiliation{Department of Condensed Matter Physics, Weizmann Institute of Science, Rehovot 7610001, Israel}

\begin{abstract}
\textbf{Weyl semimetals (WSM) have been extensively studied due to their exotic properties such as  topological surface states  and anomalous transport phenomena. Their band structure topology is usually predetermined by material parameters and can hardly be manipulated once the material is formed. Their unique transport properties appear usually at very low temperature, which sets challenges for practical device applications. In this work, we demonstrate a way to modify the band topology via a weak magnetic field in a ferromagnetic topological semimetal, Co$_2$MnAl, at room temperature.
We observe a tunable, giant anomalous Hall effect, which is induced by the transition between Weyl points and nodal rings as rotating the magnetization axis.
The anomalous Hall conductivity is as large as that of a 3D quantum anomalous Hall effect (QAHE), with the Hall angle reaching a record value (21\%) at the room temperature among magnetic conductors. Furthermore, we propose a material recipe to generate the giant anomalous Hall effect by gaping nodal rings without requiring the existence of Weyl points. Our work reveals an ideal intrinsically magnetic platform to explore the interplay between magnetic dynamics and topological physics for the development of a new generation of spintronic devices. 
}
\end{abstract}
\maketitle

\section{Introduction}
The Weyl semimetal (WSM)~\cite{Wan2011,volovik2003universe,murakami2007phase,Burkov2011,Hosur2013,Yan2017,Armitage2017} is characterized by the linear-band-crossing points, called Weyl points, which exhibits monopole-type structure of the Berry curvature~\cite{Nagaosa2010,Xiao2010}, leading to many exotic properties such as the Fermi arc surface states~\cite{Wan2011}, the chiral anomaly effect~\cite{Nielsen1981a,Nielsen1983} and the anomalous Hall effect (AHE)~\cite{Yang2011QHE,Burkov2014}. This topological phase has been discovered in materials such as TaAs~\cite{Weng2015,Huang2015,Lv2015TaAs,Xu2015TaAs,Yang2015TaAs} and MoTe$_2$\cite{soluyanov2015type,Sun2015MoTe2,Jiang2016,Deng2016} 
recently. 
In these materials, once the crystal is formed, their positions and energies of Weyl points are usually predetermined by material parameters, including the crystal structure and spin-orbit coupling, and can hardly be manipulated freely. 

A Weyl point is robust in the sense it does not require any symmetry protection except the lattice translation. Inside a mirror plane that prohibits the Berry-curvature monopole, the Weyl point disappears while 1D nodal lines may emerge due to the mirror symmetry~\cite{Chiu2014,Fang2015,Fang2016}. The nodal line displays a $\pi$ Zak-Berry phase accumulated along a loop circling the nodal line,  which induces Shockley-like~\cite{Shockley1939,Yan2015} surface states(called drum-head surface states\cite{Bian2016drumhead}). However, the nodal line exhibits zero Berry curvature in its vicinity and thus does not generate an AHE. If the mirror symmetry is broken, the nodal line gets gapped out and evolves into a pair of Weyl points(e.g. Ref. \onlinecite{Ominato2019}). The topology change can be directly probed by the AHE or observed in other topology-induced phenomena. Therefore, the rotation of the magnetization orientation, which sensitively switches the mirror symmetry in a ferromagnet, provides a powerful tool to tune the topological band structure.

Although there have been several reported/predicted  magnetic WSMs, few of them are are appropriate for tuning the band structure topology via magnetic field. For the antiferromagnetic WSMs Mn$_3$Sn and its sister compound Mn$_3$Ge\cite{yang2017topological,Nayak2016,Nakatsuji2015,Kuroda2017,Li2018Mn3Sn}, in spite of  their AHE varying with the rotation of spins within the spin-easy-plane, it remains unclear how their Weyl points and nodal lines (if they exist) evolve because of the complexity in the band structure. Co$_3$Sn$_2$S$_2$ was reported to be a layered ferromagnetic (FM) WSM~\cite{Liu2018,Wang2018,Guin2019,Ding2019,Morali2019}, which presents only Weyl points near the Fermi energy. Since its magnetization favor only the out-of-plane direction with  the Curie temperature at 177 K, the band tuning by magnetization is challenging, which is similar to the case of a recently reported FM nodal line semimetal Fe$_3$GeTe$_2$\cite{Kim2018}.
Recently,  Co-based Heusler alloy compounds Co$_2$XZ(X=V, Zr, Nb, Ti, Mn, Hf; Z =Si, Ge, Sn, Ga and Al), previously known as half metallic ferromagnets, were predicted to host FM WSM phases~\cite{Kubler2016,Chang2016b,Chang2017,Manna2018,Noky2019}. Since many of these materials are soft ferromagnets with their Curie temperatures far above room temperature, a weak external magnetic field can easily drive their magnetization to rotate in a wide temperature range. Hence these  materials provide ideal platforms to tune the band structure topology.  

Although Co$_2$XZ allows for many different element combinations, experimental studies on their possible exotic properties induced by the expected  WSM states are sparse, which is possibly due to the difficulty of the single crystal growth of this family of materials. Up-to-date,  Co$_2$MnGa is the only member which was reported to have distinct properties associated with the FM WSM state, i.e. a giant AHE and anomalous Nernst effect\cite{Sakai2018,Guin2018}. The band structure calculations show that  Co$_2$MnGa as well as  other Co$_2$XZ members have both Weyl points and nodal ring and the giant AHE in these materials is generally believed to originate from Weyl points \cite{Kubler2016,Sakai2018,Noky2019,Chang2016b}, with the anomalous Hall conductivity (AHC) being proportional to the separation to Weyl points. In theory, the largest possible Weyl point separation is the Brillouin size, which should induce the quantized AHC, i.e. a 3D quantum anomalous Hall effect (QAHE). How can a magnetic WSM reach the  3D QAHE or giant AHE ? What is the recipe to find such materials? Answers to these questions are not only of fundamental importance, but also likely leads to technological applications. 

In this letter, we give clues to these questions through the study of AHE of a Heusler alloy Co$_2$MnAl. This compound has attracted much attention due to its potential applications in spintronics\cite{Oogane2006} and Hall sensors \cite{Vilanova2011,Chen2004} as well as its recent prediction of FM WSM\cite{Kubler2016}. All previous studies on this material were based on either polycrystalline or thin film samples. Our recent success in growing Co$_2$MnAl single crystals has enabled us to observe its fascinating properties originating from the band topology.
We find this material exhibits very large AHC, up to 1300 $\Omega^{-1}$cm$^{-1}$ at room temperature; more noticeably, its room temperature anomalous Hall angle reaches a new record value 21 \% among all magnetic conductors. Furthermore, its AHE can be tuned by rotation of the magnetization axis as expected. From the comparison between experimental results and theoretical calculations, we find the gapped nodal rings induce large Berry curvature, which is responsible for the observed giant AHE.  Our findings demonstrate a new mechanism of creating a large AHE by gaping nodal rings; through this mechanism, an AHC as large as that of a 3D QAHE can be reached.  This mechanism is distinct from the previous understanding of the AHE originating from Weyl points  for Co$_2$MnGa and similar materials. 

\section{Results}
\subsection{Experiment}

  \begin{figure*}
  \centering
      \includegraphics[width=\linewidth]{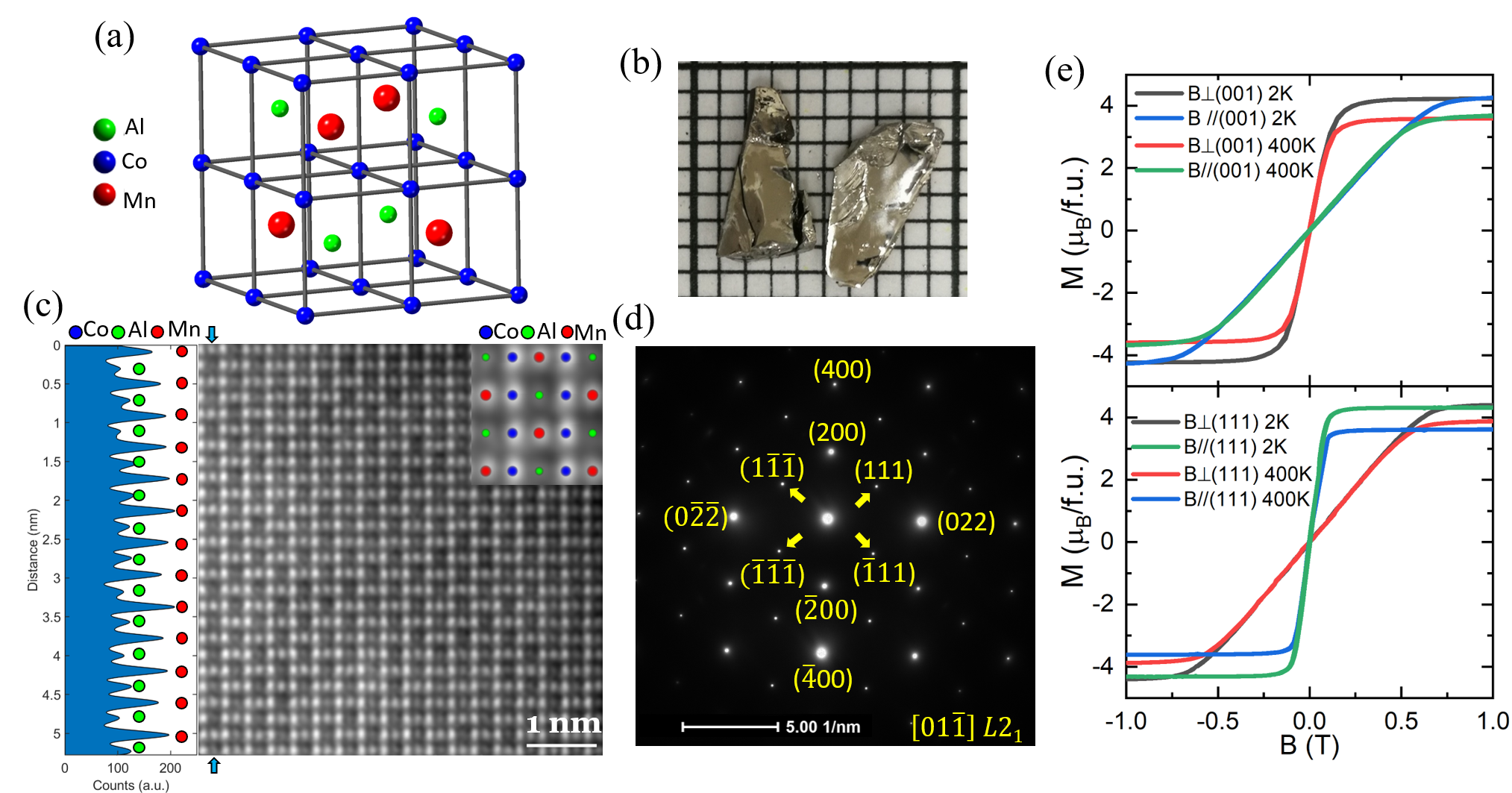}
    \caption{(a) Schematic crystal structure of L2$_1$ type Co$_2$MnAl, b) the optical graph of Co$_2$MnAl single crystals grown by the optical floating zone method; (c) HAADF-STEM image of a selected area taken along the [110] zone axis. The left panel shows alternating intensity due to Z-contrast, revealing the ordering of Mn and Al; the right inset presents a magniﬁed image with the atoms overlaid on top. (d) The  selected area diffraction pattern taken along the [110] zone axis. (e) The isothermal magnetization data at 2 K and 400 K, measured with the magnetic field applied parallel and perpendicular to the (001)/(111) plane respectively.}
    \label{Fig:crystal}
\end{figure*}

The Co$_2$MnAl single crystals used in this study were grown using floating-zone technique  (see the supplementary material (SM)). Figure 1b  shows an optical image of typical crystals. Our powder X-ray diffraction measurements confirmed its cubic structure. Since Co$_2$MnAl can have several structural phases with different disorder types, including L2$_1$ (Mn, Al ordered), B2 (Mn, Al disordered), DO$_3$ (Co, Mn disordered), A2 (Co, Mn, Al disordered)~\cite{Vilanova2011} and only the L2$_1$ phase with the cubic space group of $Fm\overline{3}m$ was predicted to host the FM WSM phase, we have performed careful structural analyses using transmission electron microscopy to find if our samples possess the L2$_1$ phase (see the SM for experimental details). As shown in Fig. 1a, in the L2$_1$ phase, Mn and Al are ordered, which is manifested by the (111) diffraction spot/peak in an electron/X-ray diffraction pattern according to previous studies\cite{Umetsu2008}. From our electron diffraction analyses, we have indeed observed the (111) diffraction spot in the electron diffraction pattern taken along the [110] zone-axis (Fig. 1d), indicating our Co$_2$MnAl crystals surely have the L2$_1$ structure phase. This is further corroborated by scanning transmission electron microscopy (STEM) imaging shown in Fig. 1c where the periodic, alternating distribution of Mn and Al (left inset to Fig. 1b) can be seen clearly from the atomic intensity line due to Z-contrast.  Furthermore, we have also used the High Angle Annular Dark Field STEM technology to check the sample homogeneity, which confirmed that uniform L2$_1$ phase is formed throughout the entire sample. The magnetic properties of Co$_2$MnAl single crystals were characterized through magnetization measurements.  Fig. 1e shows the isothermal magnetization data at 2 K and 400 K, measured with the magnetic field applied parallel and perpendicular to the (001)/(111) plane respectively. These data are consistent with the previous report that Co$_2$MnAl is a soft ferromagnet, with the saturation moment Ms of about 4.2 $\pm$ 0.2 $\mu_B$ per formula \cite{Umetsu2008}. The small decrease of Ms from 2 K to 400 K suggests that its Curie temperature is far above room temperature, which can not be probed in our SQUID magnetometer. The previously-reported $T_c$ for this material is 726 K\cite{Umetsu2008}. Moreover, our magnetization data also reveals its FM properties are anisotropic and the (111) plane is the spin easy plane.

  \begin{figure*}[t]
  \centering
      \includegraphics[width=\linewidth]{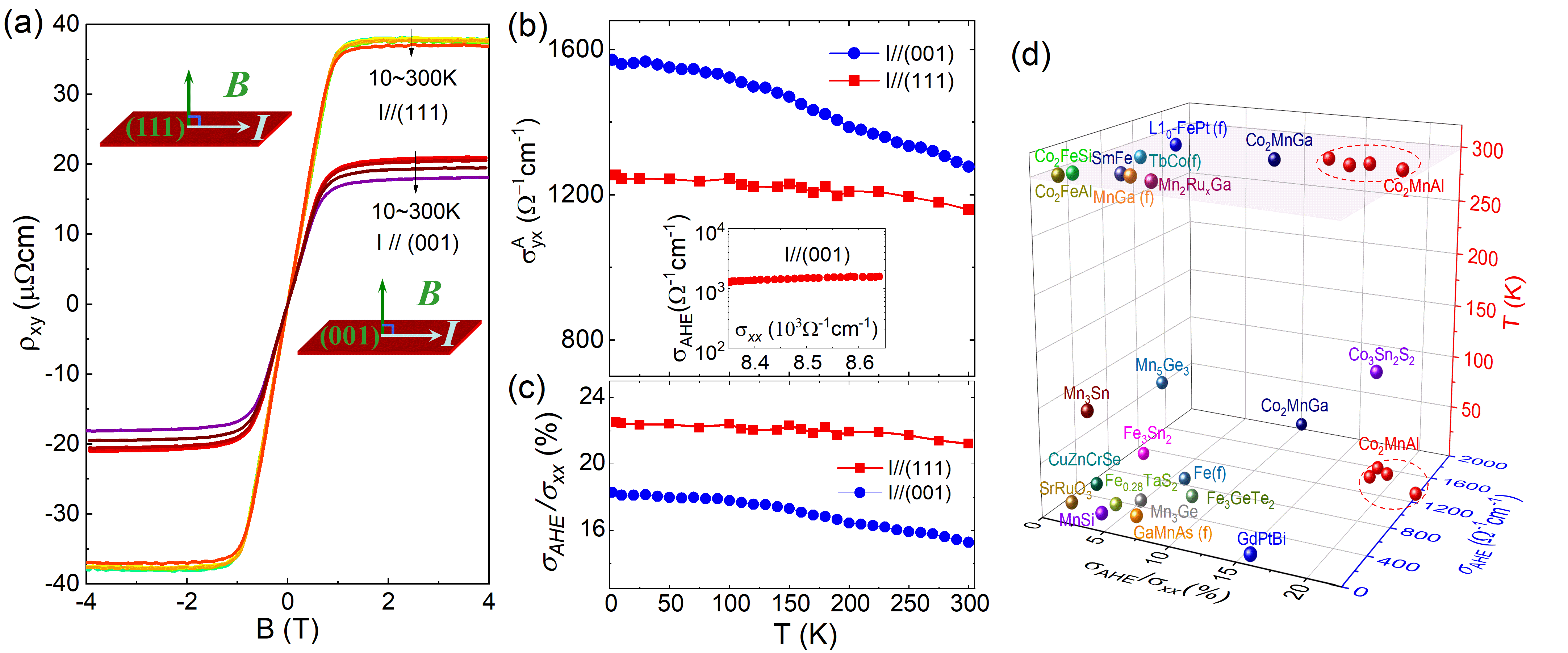}
    \caption{(Color online) (a) Magnetic field dependent Hall resistivity ($\rho_{xy}$) at different temperatures for two Co$_2$MnAl samples with different orientations. The insets show the Hall measurement set-up for these two samples, with I//(001) and I//(111).  (b) Temperature dependence of  Anomalous Hall conductivity (AHC) ($\sigma^A_{yx}$) for the samples shown in Fig. 2a. The inset illustrates the anomalous Hall conductivity as a function of longitudinal conductivity ($\sigma_{xx}$), which is derived from the  $\rho_{xy}$ and  $\rho_{xx}$ values measured at 2T. (c)The temperature dependence of Hall angle (AHA) $\sigma^{A}_{yx}/\sigma_{xx}$ at 2 T. (d) Comparison of AHC and AHA between Co$_2$MnAl and other magnetic conductors;``(f)'' refers to thin-film materials. Co$_2$MnAl exhibits both large AHA and large AHC in a wide temperature range. In addition to the data collected on the samples shown in Fig. 2a, we have also included the data measured on another two additional  Co$_2$MnAl samples with $I$//(001) and $I$//(010) in this figure. The AHC and AHA data of other cited magnetic materials and the related references are given in Fig.~\ref{Fig:TableS1} in SM.}
        \label{Fig:Hall}
\end{figure*}

We have performed Hall resistivity $\rho_{xy}$ measurements on the  Co$_2$MnAl samples with the electric current applied to the (001), (010)  and (111) planes respectively. Figure 2a presents the field dependence of $\rho_{xy}$ in the 10-300 K temperature ranges for two typical samples with $I$//(001) and $I$//(111). $\rho_{xy}$ shows nearly the same magnetic field dependence as the isothermal magnetization curve shown in Fig.1e, indicating that the anomalous contribution (proportional to $M$) to the Hall effect is dominant and the normal contribution due to the Lorentz effect (proportional to $B$) is negligibly small at 300 K. As seen in Fig. 2a, $\rho_{xy}$ at room temperature show large values, about 15~$\mu \Omega$cm for $I$//(001) and 37 $\mu \Omega$cm for $I$//(111). We have also measured longitudinal resistivity $\rho_{xx}$ for these samples (data not shown here). From the $\rho_{xy}$ and $\rho_{xx}$ data, we have derived the anomalous Hall conductivity (AHC) $\sigma^A_{yx}$ as a function of temperature at 2T for the $I$//(001) and $I$//(111) samples (Fig. 2b) via tensor conversion, i.e. $\sigma^A_{yx}$ = $\rho_{xy}$/ ($\rho_{xy}^2$ +$\rho_{xx}^2$).  
Both samples are found to show exceptionally large AHC. At 2K, $\sigma^A_{yx}$ reaches 1600 $\Omega ^{-1}$cm$^{-1}$ for the I//(001) sample. Upon increasing temperature, the AHC decreases slightly and remains up to 1300 $\Omega ^{-1}$cm$^{-1}$ at room temperature. For the $I$//(111) sample, the AHC is almost temperature-independent and just decreases from 1250 $\Omega ^{-1}$cm$^{-1}$ at 2 K to 1190 $\Omega ^{-1}$cm$^{-1}$ at room temperature. Such a large AHC at room temperature in Co$_2$MnAl has never been reported due to the lack of bulk single crystals. We noticed that Co$_2$MnAl thin films were previously made using radio frequency (RF) magnetron sputtering \cite{Vilanova2011}. Its room temperature $\rho_{xy}$ is about 20 $\mu \Omega$cm, comparable to the $\rho_{xy}$ value seen in our I//(001) sample. However, ref. \onlinecite{Vilanova2011} did not report $\rho_{xx}$ or $\sigma^A_{yx}$ data for Co$_2$MnAl films. The giant room temperature  AHC revealed in our experiment, as well as its tunability by a weak magnetic field as demonstrated below, suggest Co$_2$MnAl is a promising material for device applications such as Hall sensor.  

Besides large AHC, Co$_2$MnAl also shows a large anomalous Hall angle (AHA) (defined as $\Theta^H=\sigma^{A}_{yx}/\sigma_{xx}$),  as shown in Fig. 2c. For the $I$//(111) sample,  its $\Theta^H$ value is as large as 21\% even at room temperature, which is a record value among either trivial or topological magnetic conductors, as far as we know. This can be seen in  In Fig. 2d, where we have compared AHC and AHA between our Co$_2$MnAl samples and those magnetic materials known as having large $\sigma_{xx}$ and $\Theta^H$. Magnetic WSMs are generally expected to have large AHC and AHA due to the Berry curvature induced by non-trivial band topology. Among reported magnetic WSMs, while some of them indeed exhibit large AHC and AHA, they occur mostly at low temperatures. For instance, the AHA of the FM WSM Co$_3$Sn$_2$S$_2$ can also reach ~20\%, but it can be observed only below 120K. The magnetic field induced WSM GdPtBi was also reported to have large AHA, with the largest value of about 10\%  being probed only below 10 K. As noted above, Heusler compound Co$_2$MnGa has also been reported to have very large AHE; its AHC reaches about 2000 $\Omega ^{-1}$cm$^{-1}$ at 2 K, but decreases down to ~1000 $\Omega ^{-1}$cm$^{-1}$ at room temperature. Its room-temperature AHA is - 12\%, about half of the largest value we observed in Co$_2$MnAl. Compared to Co$_2$MnGa, Co$_2$MnAl shows a much weaker temperature dependence in AHC and AHA for the $I$//(111) plane (Fig. 2c), which is important for device applications.

In general, an AHE may either originate from extrinsic mechanism (i.e skew scattering and side jump), or arises from intrinsic Berry curvature contribution (i.e. the effective filed in momentum space created by the non-trivial band topology), or combined extrinsic and intrinsic contributions. The large AHC revealed in our experiments for Co$_2$MnAl shows clear signatures consistent with intrinsic AHE. In the inset of Fig.2(b), we present $\sigma^A_{yx}$ as a function of $\sigma_{xx}$ [ $\sigma_{xx} = \rho_{xx}/(\rho_{xy}^2 +\rho_{xx}^2)$] in the 2 - 300 K temperature range for the I//(001) sample. $\sigma_{xx}$ varies  in the range of 8350 - 8650 $\Omega ^{-1}$cm$^{-1}$, which resides in the moderately dirty region, where the AHE should be attributed to the intrinsic Berry curvature origin according to the theory of AHE in ferromagnets \cite{Xiao2010}. Moreover, from the inset to Fig. 2b, it can be seen that $\sigma_{yx}$ hardly varies with $\sigma_{xx}$, which is a commonly expected  scaling behavior for an intrinsic AHE \cite{Nagaosa2010} in an intermediate regime between the good metal and bad metal. In the I//(111) sample, we observed a similar behavior.  

\begin{figure*}
\includegraphics[width=\linewidth]{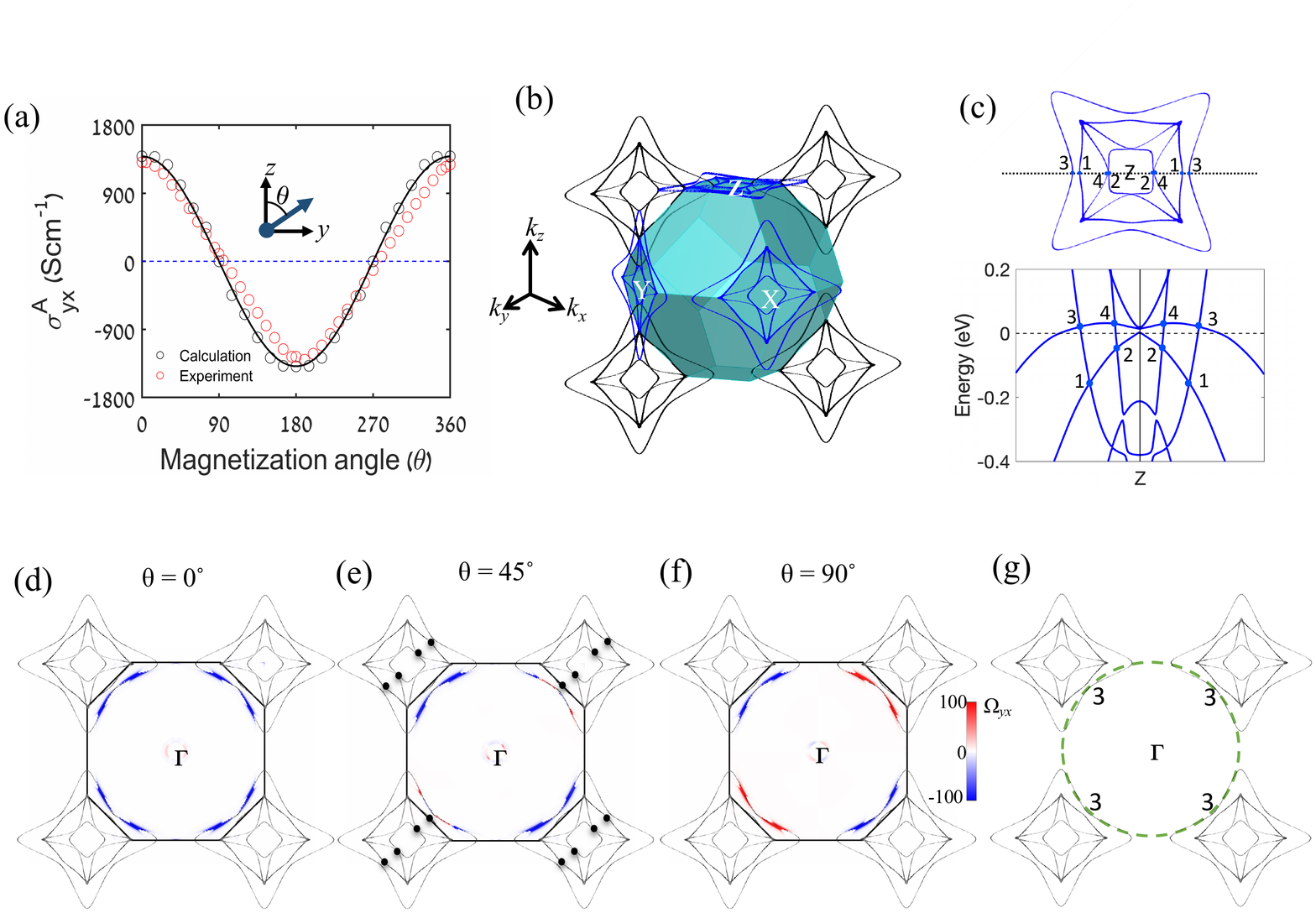}
\caption{\textbf{The anomalous Hall effect and topological band structure.}
(a) The anomalous Hall conductivity $\sigma^A_{yx}$ as rotating the magnetization from the $z$ to $y$ axis. The experimental and theoretical results are represented by red and black circles, respectively. The solid curve is a cosine profile to guide eyes. 
(b) Nodal rings and the first Brillouin zone of Co$_2$MnAl. Without SOC, there are nodal rings in these mirror planes, $k_{x,y,z}=0$ or $\pi$. For example, there are four nodal rings that are labeled as \#1-4 in (c), centered at the $Z$ point of the FCC Brillouin zone. When magnetization align along the [001] direction by SOC, all nodal rings are gapped out because of the mirror symmetry breaking, except those in the $k_{z}=0$ or $\pi$ plane. The preserved nodal rings and related band structure are shown in (c). The band dispersion, in which the nodal ring crossing points are indicated by blue dots, is along the black dotted line in the upper panel.
(d)-(f) The Berry curvature ($\Omega_{yx}$) distribution in the $k_x =0$ plane centered at $\Gamma$ for different magnetization angle ($\theta$). Nodal rings (black) are gapped out in this plane.  Near the charge neutral point, the Berry curvature is mainly contributed by the gapped nodal ring \#3. When $\theta=45^{\circ}$, nodal rings \#1 \& 4 evolve into Weyl points (filled black circles in (e)) while nodal rings \#2 \& 3 are fully gapped. (g) The nodal rings \#3 at four corners can be treated as the reconstruction of a large ring, which is illustrated by a green dashed circle, centered at $\Gamma$.}
\label{Fig:sigma}
\end{figure*}

\subsection{Discussion}

To understand the intrinsic AHE of Co$_2$MnAl, we have preformed theoretical calculations on the band structure and AHC of Co$_2$MnAl. The intrinsic AHE originates from the band structure and is robust against disorders and defects. The magnetization direction can modify the AHE, because it determines the symmetry and consequently the band structure of the system. We show the calculated AHC ($\sigma^A_{yx}$) in Fig. \ref{Fig:AHC_tem_angle}. It is not surprising to find that the magnetization along [001] and [111] leads to slightly different amplitude in $\sigma^A_{yx}$. Near the charge neutral point, the [111] value is slightly lower than the [001] value and the [111] AHC shows weaker temperature dependence than the [001] case, which is semi-quantitatively consistent with our experiment and is attributed to the band structure anisotropy.

We examine the origin of the strong AHE in the band structure by taking the [001] magnetization for example. In an energy window of $-0.2$ to $+0.1$ eV with respect to the charge neutral point ($\mu=0$), we find four nodal rings centered at the $Z$ point of the FCC Brillouin zone (see more information in the SM). We note that these nodal rings are protected by the mirror symmetry of the $k_z=0$ ($k_z=0$ and $\pi$ planes are equivalent in the FCC Brillouin zone), as shown in Figs.~\ref{Fig:sigma}c and ~\ref{Fig:nodalBerry}b. However, corresponding nodal rings at the $k_x=0$ and $k_y=0$ planes are gapped out because of the mirror symmetry breaking caused by the [001] magnetic moment (as shown in Fig.~\ref{Fig:nodalBerry}c). These four nodal rings, which are denoted by \#1-4 in the following discussions, are induced by band crossing between the highest two valence bands and lowest two conduction bands (see Fig.~\ref{Fig:sigma}c). In previous study (Ref. \onlinecite{Chang2017}), only nodal rings \#2-3 were investigated in a similar compound Co$_2$MnGa without considering the SOC. Without SOC, the magnetization axis does not pick up a specific direction. Thus, these nodal rings appear in all mirror planes, i.e. $k_{x,y,z}=0$. As shown in Fig.~\ref{Fig:sigma}b, four nodal rings in different planes interconnect each other to form topological Hopf nodal links\cite{Yan2017nodal,Chang2017}.
The SOC leads the magnetic moment to couple to the lattice, thus reducing the symmetry. The preserved nodal rings in the mirror plane do not contribute any Berry curvature (see Fig.~\ref{Fig:nodalBerry}), which is characterized by a $\pi$ Berry phase accumulated along a loop interlocking the nodal ring. Thus, these four nodal rings do not contribute to the AHE. In contrast, the gaped nodal rings generate large Berry curvature and thus give rise to the huge AHC (see Fig.~\ref{Fig:nodalBerry}) as observed in our experiments. Four peaks of $\sigma^A_{yx}$ in this energy window are dominantly contributed by these four gaped nodal rings (see Fig.~\ref{Fig:AHC_Fermi}). Near the charge neutral point, AHC exhibits a large value because of the existing peak at $\mu=25$ meV, which is mainly induced by the gaped nodal ring \#3 (see Fig.~\ref{Fig:AHC_Fermi}). 

We can control the AHE response by rotating the magnetization axis, when fixing the current and Hall voltage probes inside the (001) plane. In experiment, we observed a $\cos(\theta)$-like oscillation of $\sigma^A_{yx}$ with respect to the magnetization angle, $\theta$. As shown in Fig. \ref{Fig:sigma}a, the experiment is well consistent with our theoretical calculations. As the magnetization changes from [001] to [010], nodal rings appear in the the $k_y=0$ plane while they disappear in $k_{x,z}=0$ planes. In between [001] and [010], \#3 nodal rings are fully gaped in all three planes, though some other nodal rings (e.g. \#1 and \#4) evolve into Weyl points (Fig.~\ref{Fig:sigma}e). As a consequence, the induced Berry curvature $\Omega_{yx}$ evolves dramatically from [001] to [010], as shown in Fig. \ref{Fig:sigma}d-f, leading to the oscillating $\sigma^A_{yx}$ as seen in experiments. The nearly cosine-like profile is consistent with the fact that Co$_2$MnAl exhibits weak anisotropy in the magnetization orientation in experiment. We should point out that present $\cos(\theta)$-type AHE is different from the known planar Hall effect, which has a $sin(2\theta)$ dependence\cite{Ky1968}. The planar Hall effect appears in a FM metal where the magnetization remains unchanged as rotating an in-plane magnetic field, or a nonmagnetic WSM where the chiral anomaly plays a role\cite{Burkov2017,Nandy2017}.

\begin{figure}[t]
\centering
\includegraphics[width=\linewidth]{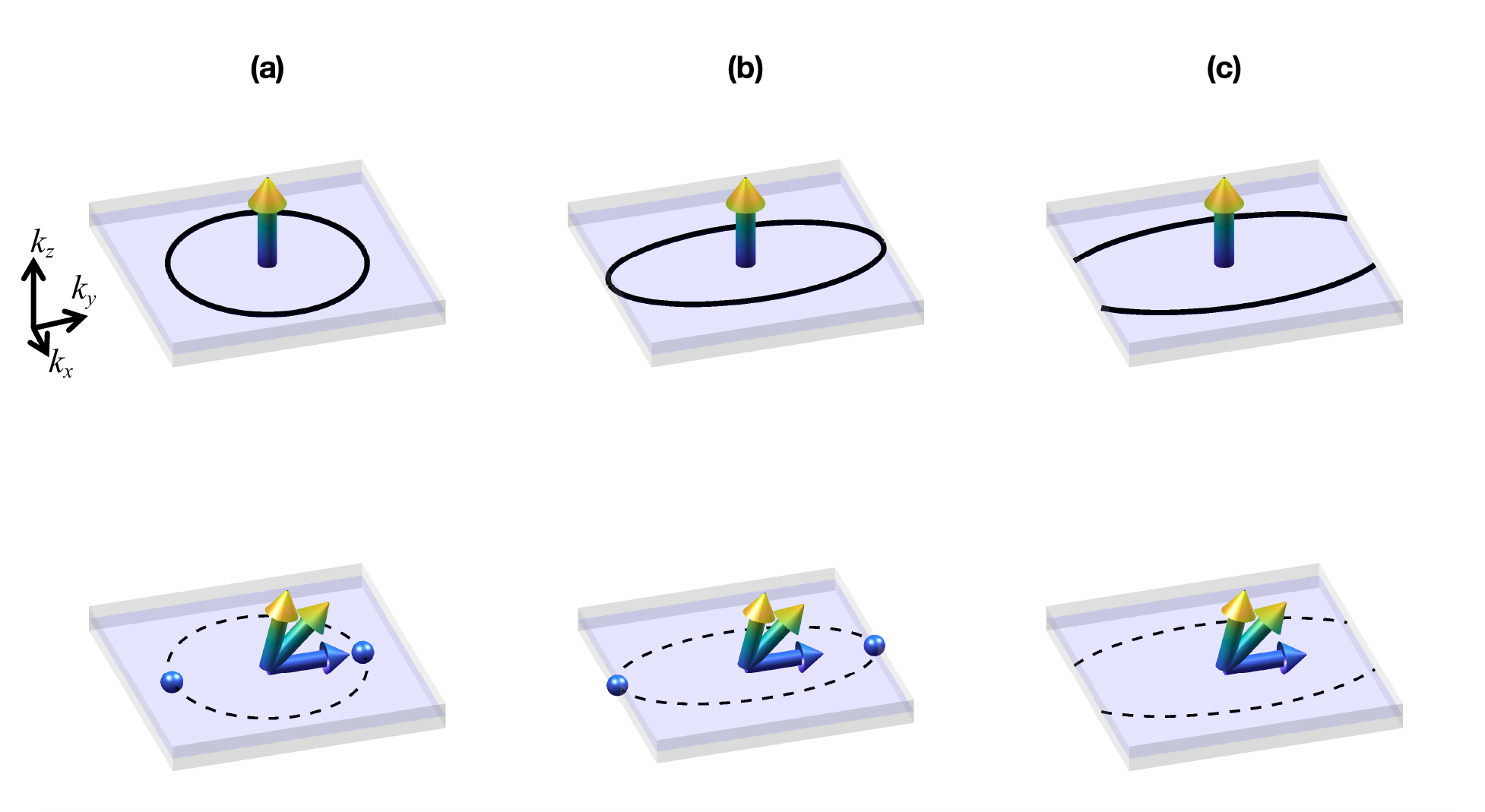}
\caption{\textbf{Schematic transition between nodal rings and Weyl points.} (a) The nodal ring (solid ring) is protected by the in-plan mirror symmetry where the spin (arrow) points out of plane, in the upper panel.  If the spin rotates to break the mirror symmetry, the nodal ring gets gapped out (dashed ring), giving rise to a pair of Weyl points (blue spheres). The induced AHE is proportional to the separation of the Weyl points, i.e. the diameter of the nodal ring ($k_d$) in a form $\frac{e^2}{h} \frac{k_d}{2\pi a}$, where $a$ is the lattice parameter. (b) The nodal ring is as large as the Brillouin zone size, as a critical point. The resultant Weyl point is pushed to the zone boundary and meets another Weyl point with opposite chirality from the second zone. The induced AHE is quantized to $\frac{e^2}{h} \frac{1}{a}$.   (c) The nodal ring is larger than the Brillouin zone size, as an open nodal ring. Then the gapped ring does not induce Weyl points in the open direction, generating a 3D quantized Hall conductivity, $\frac{e^2}{h} \frac{1}{a}$. }
\label{Fig:nodalring}
\end{figure}

In principle, a simple nodal ring can evolve into a pair of Weyl points when the magnetization breaks corresponding mirror symmetry, as illustrated in Fig. 4(a). The induced AHC is proportional to the Weyl point separation~\cite{Yang2011QHE,Burkov2014}. This scenario is seemingly consistent with our observation on nodal rings \#1 and \#4. However, we do not observe the existence of Weyl points for nodal ring \#3, which is the main source of the AHE observed in experiment, when rotating the magnetization axis. 

Actually the missing of Weyl points for nodal ring \#3 can explain the large magnitude of the AHC observed in Co$_2$MnAl. Take the [001] magnetization as an example. The theoretical value of $\sigma^A_{yx} \sim 1400$ $\Omega^{-1} cm^{-1}$ is as large as the AHC of a 3D QAHE, $\frac{2e^2}{h}\frac{1}{a} = 1347 ~\Omega^{-1} cm^{-1}$ where $a$ takes the experimental value 5.75 \AA. We note that this is not coincidence. Although this nodal ring is centered at the $Z$ point, it emerges from a large nodal ring that is centered at $\Gamma$. Because it even crosses the Brillouin zone boundary, the $\Gamma$-centered ring gets reconstructed to smaller rings at the zone boundary (see Fig. \ref{Fig:sigma}g). Although smaller nodal rings generate Weyl points, a reconstructed nodal ring does not generate Weyl points (illustrated in Fig. 4c). This is because that a Weyl point, if it exists, must annihilate with another one from the neighboring Brillouin zone.
As a consequence, the band structure has a gap and the induced AHC has a quantized value $\frac{e^2}{h}\frac{1}{a}$. In our material, there are two groups of gapped nodal rings on the $k_{x,y}=0$ planes. Therefore, the ideal AHC is $\frac{2e^2}{h}\frac{1}{a}$. Because the nodal ring cannot be fully gapped out in the Fermi energy and there are also contribution for other bands, we do not expect a true 3D QAHE here. We note that the same scenario can also be applied to Co$_2$MnGa.
In addition, if the $\Gamma$-centered ring was smaller than the Brillouin zone size in Co$_2$MnAl, it could generate Weyl points but smaller AHC than the quantized value. This is actually the case of Heusler compounds Fe$_2$Mn$X$ ($X =$ P, As, Sb)\cite{Noky2019}, Co$_2$VGa~\cite{Manna2018}, and Co$_2$Ti$X$ ($X$=Si, Ge, Sn)\cite{Chang2016b}. We can rationalize these full Heusler magnets in the same framework here.

Based on the understanding of Co$_2$MnAl, we can obtain useful insights to design magnetic materials with strong AHE. To generate large AHC, we need multiple large nodal rings that are larger than the Brillouin zone size. To host multiple nodal rings (in the absence of SOC), we prefer to have multiple mirror planes in the crystal structure, which usually indicates high-index space groups. In general, highly-symmetric magnetic materials with strong SOC are an optimal choice, not only for the AHE but also for other Berry-curvature-induced phenomena like the anomalous thermal Hall effect and anomalous Nernst effect. Similarly, nonmagnetic materials with many mirror planes and strong SOC are ideal for the spin Hall effect, as pointed out in Ref.~\onlinecite{Sun2016}.

Our theory and experiment together also demonstrate that the form of the gapless nodes (Weyl points or nodal rings), as well as their locations and velocities, all strongly rely on the magnetization axis. Therefore, the magnetization can potentially be utilized to engineer an artificial gauge field \cite{Liu2013,Hutasoit2014} and spacetime geometric structure \cite{Weststrom2017,Liang2019} of Weyl fermions. On the other hand, Weyl fermions can also mediate a strong coupling between magnetization dynamics and electromagnetic fields\cite{Hutasoit2014}, thus allowing for the electric control of magnetic dynamics in this system. Thus, Co$_2$MnAl provides an ideal intrinsically magnetic platform to explore the interplay between magnetic dynamics and topological physics for the development of a new generation of spintronic devices.

\section{Acknowledgment}
The experimental work at Tulane and Penn State is supported by the US National Science Foundation under grant DMR1707502. LM., L. M. and N.A.'s work is supported by the Penn State Center for Nanoscale Science, an NSF MRSEC under the grant number DMR-1420620.
B.Y. acknowledges the financial support by 
the Willner Family Leadership Institute for the Weizmann Institute of Science,
the Benoziyo Endowment Fund for the Advancement of Science, 
Ruth and Herman Albert Scholars Program for New Scientists, and the European Research Council (ERC grant No. 815869). C.X.L. acknowledges the support of the Office of Naval Research (Grant No. N00014-18-1-2793), the U.S. Department of Energy (Grant No.~DESC0019064) and Kaufman New Initiative research grant KA2018-98553 of the Pittsburgh Foundation.

%

\clearpage
\renewcommand{\thesection}{S\arabic{section}}
\renewcommand{\thetable}{S\arabic{table}}
\renewcommand{\thefigure}{S\arabic{figure}}
\renewcommand{\theequation}{S\arabic{equation}}

\setcounter{section}{0}
\setcounter{figure}{0}
\setcounter{table}{0}
\setcounter{equation}{0}

{\large\bf Supplementary Materials}

\section{Calculation method}
  We have performed density-functional theory (DFT) calculations with a full-potential local-orbital minimum-basis (FPLO)\cite{koepernik1999full} code to calculate electronic structure. The exchange and correlation energies are considered in the generalized gradient approximation(GGA) by Perdew-Burke-Enzerhof scheme\cite{perdew1996}. Spin-orbit coupling is included. From the DFT electronic structure, we have projected the Bloch wave function into Wannier functions (Co-3d, Mn-3d and Al-3p orbitals), to construct an effective Hamiltonian ($\hat{H}$) for the bulk material.

We have evaluated the AHC ($\sigma^A_{yx}$) and Berry curvature ($\Omega_{yx}$) by the Kubo-formula approach in the linear response scheme\cite{Xiao2010},
 \begin{equation}
  \label{sigma}
    \sigma^A_{yx}(\mu) = -\frac{e^2}{\hbar}\int d\xi \frac{\partial f (\xi-\mu)}{-\partial \xi}
    \int_{BZ} \frac{d\textbf k}{(2\pi)^3}\sum_{\epsilon_n<\mu}\Omega^n_{yx}(\textbf k)   \end{equation}
    \begin{equation}
    \Omega_{yx}^n(k) = i \sum_{m\ne n } \frac{\langle n|\hat{v}_y|m \rangle \langle n|\hat{v}_x|m \rangle - (i\leftrightarrow j) }{(\epsilon_n\textbf{(k)} - \epsilon_m\textbf{(k)})^2}.
  \end{equation}
Here $\epsilon_n$ is the eigenvalue of the $|n\rangle$ eigenstate, and $\hat{v}_i=\frac{d\hat{H}}{\hbar dk_i}$ ($i=x,y,z$)  is the velocity operator, $\mu$ is Fermi level of the system, $f (\xi-\mu)$ is the Fermi-Dirac distribution. A $k$-point of grid of $300\times300\times300$ is used for the numerical integration in Equation~\ref{sigma}.

  \section{Detailed calculation result of magnetization along [001]}
 Here we explain more detailed results of the [001] magnetization. 
 The top two valence bands and bottom two conduction bands cross each other, resulting in four crossing points, as shown in Fig.~\ref{Fig:nodalBerry} (b). The equivalence of FCC Brillouin zone at the $k_z=0$ and $k_z=\pi$ planes can be found in the Fig. \ref{Fig:nodalBerry} (a). As the result of the magnetization, $M_x$ or $M_y$ mirror symmetry is broken and the nodal rings in the $k_{x,y}=0$ plane are gapped out and produce huge Berry curvature. Meanwhile, nodal rings in the $k_z=0$ plane are preserved by the $M_z$ symmetry and generate zero Berry curvature. Nodal rings in different planes are inter-connected to each other. For example, in Fig. \ref{Fig:sigma} (c), nodal rings at the $Z$ plane is locked to nodal rings in the $X$ plane from another Brillouin zone.  
 
  Since the Nodal lines has energy dispersion, we can find four peaks near Fermi energy relate with energy dispersion (Fig.\ref{Fig:AHC_Fermi}). Peak A is from gapped nodal ring \#1, Peak B and C are from gapped nodal rings \#2-3 and peak D from gapped nodal ring \#4. The peaks arise when the Fermi energy shift cross the SOC gap of each gapped nodal rings. In addition, we also check spatial Berry curvature contribution in the first Brillouin zone. As expected, nothing can be found far from the nodal rings regions.

 \begin{figure}
\centering
\includegraphics[width=\linewidth]{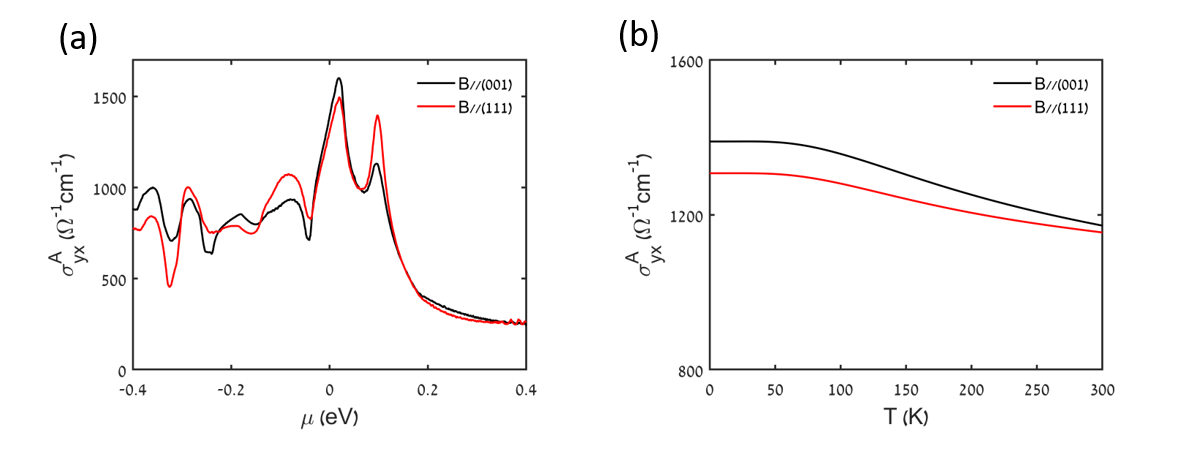}
\caption{\textbf{Compare the [001] and [111] directions.} Calculated anomalous Hall conductivity with respect to the chemical potential (a) and the temperature (b) for the magnetization along [100] (black) and [111] (red) directions.}
\label{Fig:AHC_tem_angle}
\end{figure}

\begin{figure}
\centering
\includegraphics[width=\linewidth]{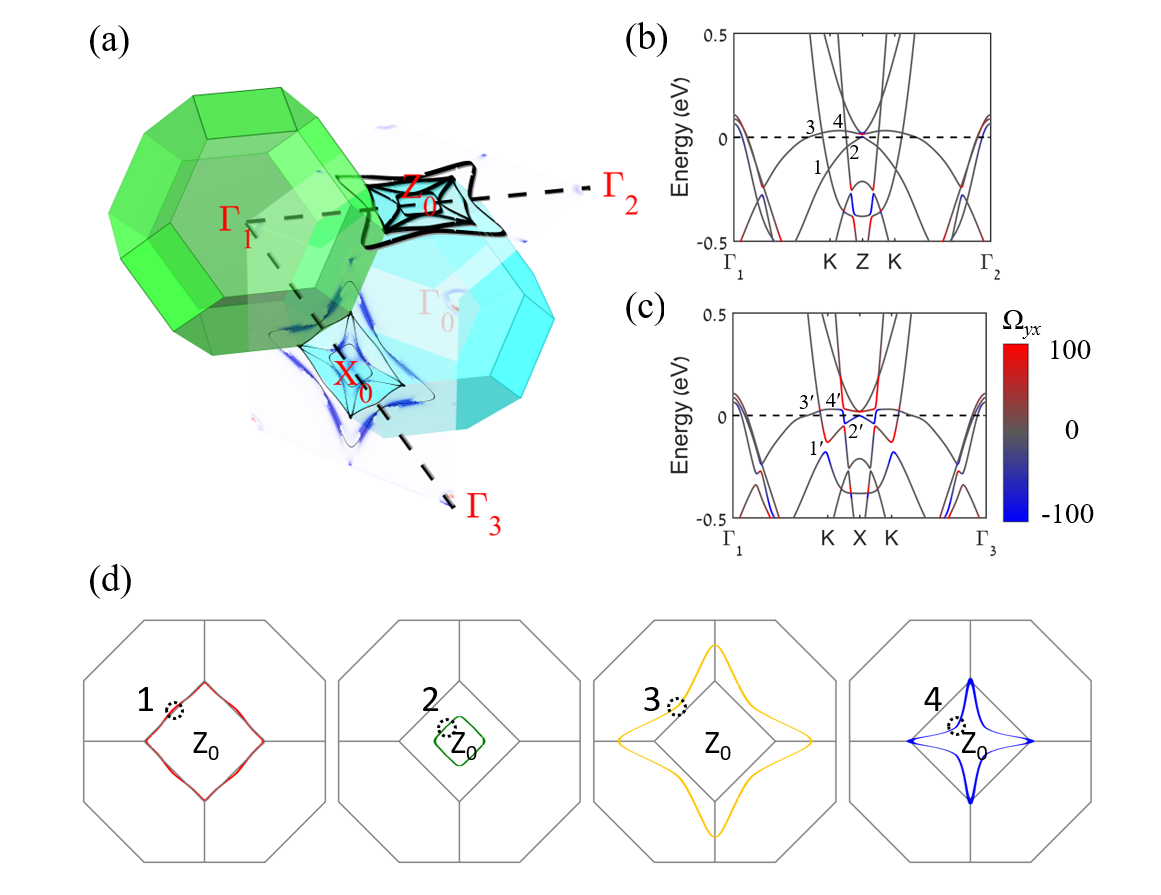}
\caption{\textbf{Band structure, Berry curvature and nodal rings.} (a) The Berry curvature at charge neutral point with gapped nodal rings in the $X_0$ plane and gappless nodal rings in the $Z_0$ plane. The magnetization axis is along the [001] direction. Green FCC Brillouin zone is the second Brillouin zone to the first one (blue). (b) Calculated band structure with Berry curvature from $\Gamma_1$ to $\Gamma_2$ as shown by the dashed line in (a). We label four crossing-points near the Fermi energy by \#1-4. They are part of the nodal rings in the $Z_0$ plane and protected by the (001) mirror symmetry. These nodal rings does not generate the Berry curvature. (c) Calculated band structure with Berry curvature from $\Gamma_1$ to $\Gamma_3$ as shown by the dashed line in (a). All four cross-points (noted as \# $1^\prime - 4^\prime $) are gapped and produce huge Berry curvature. (d) Each nodal ring in the $Z_0$ plane.}
\label{Fig:nodalBerry}
\end{figure}

\begin{figure}
\centering
\includegraphics[width=\linewidth]{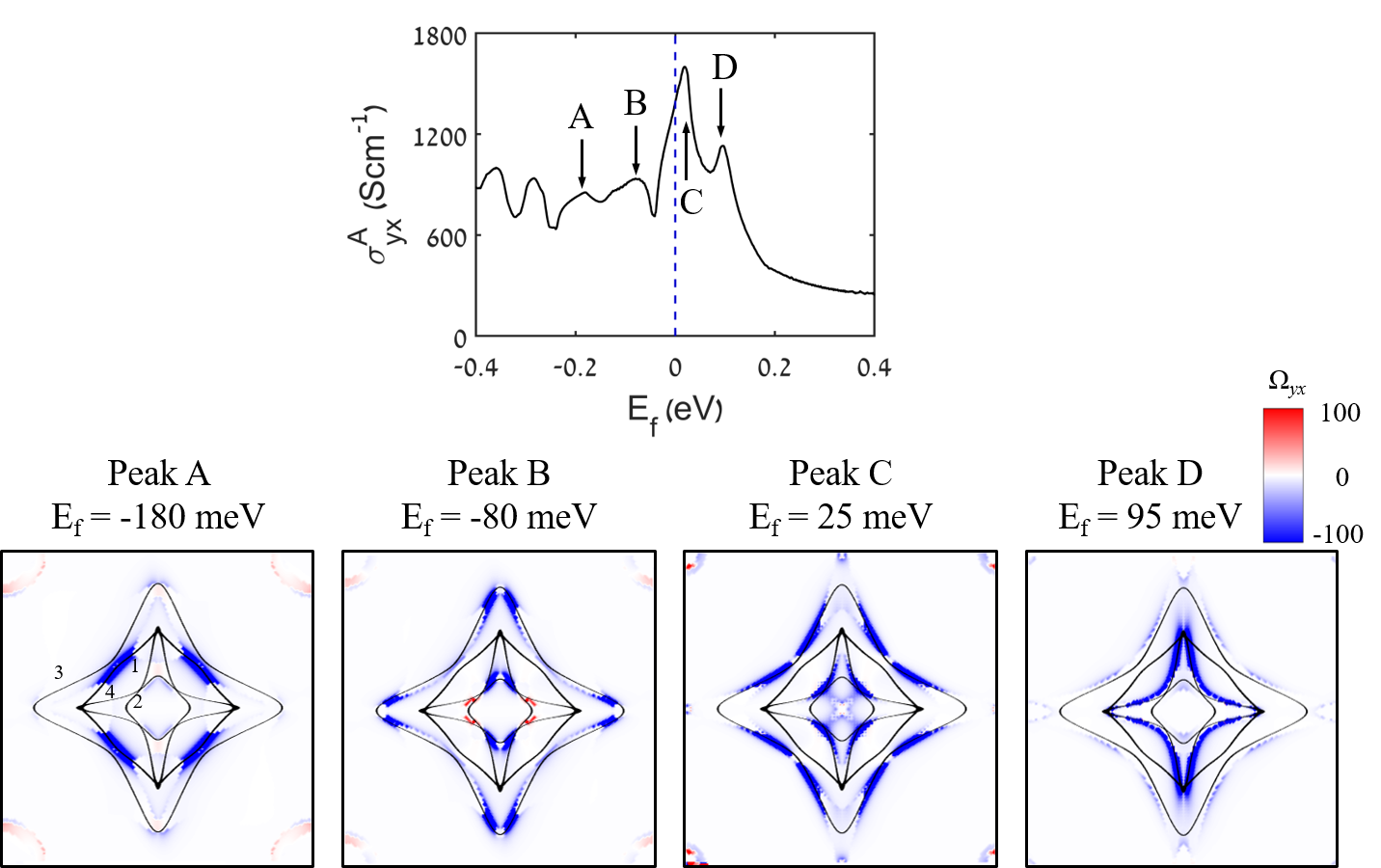}
\caption{\textbf{ The anomalous Hall conductivity ($\sigma_{yx}^A$) and nodal rings.} The calculated anomalous Hall conductivity with respect to the Fermi energy. The magnetization axis is along [001].
The charge neutral point is set to zero. Near the charge neutral point, there are four peaks of $\sigma_{yx}^A$ noted as A,B,C and D. 
We show the gapped nodal rings in the $k_x=0$ plane (see Fig.~\ref{Fig:sigma}) with the Berry curvature ($\Omega_{yx}$) in the low panels. The $\sigma_{yx}^A$ is mainly contributed by the gapped nodal rings. The peak C, which is the most relevant to the experiment, is induced mainly by the ring \#3 and also slightly by the ring \#2. Peak A is due to nodal ring \#1, peak B due to \#2 and slightly also \#3 and peak D due to \#4.}
\label{Fig:AHC_Fermi}
\end{figure}

\section{Experimental methods}

 \begin{figure}
\centering
\includegraphics[width=\linewidth]{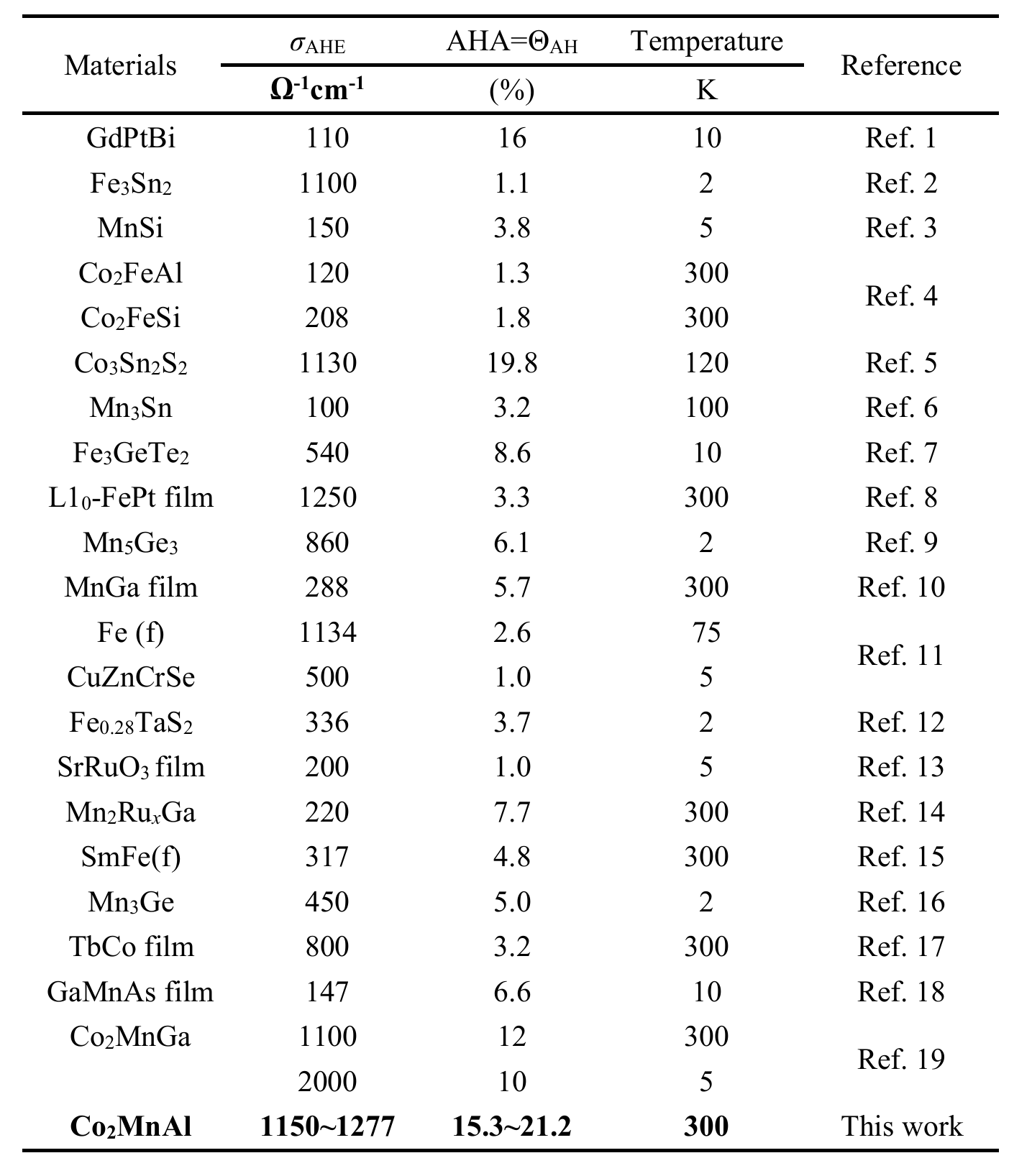}
\caption{\textbf{Summary of anomalous Hall conductivity ($\sigma_{AHE}$) and anomalous Hall angle (AHA) of currently studied magnetic materials.} Some AHA values are calculated by using the data from the references. Only the maximum values of AHA are adopted here. For Co$_2$MnGa and Co$_2$MnAl, the largest values at low temperature are also presented for comparison.}
\label{Fig:TableS1}
\end{figure}

Co$_2$MnAl single crystals were grown using floating zone (FZ) technique. The material rods used for the FZ growth were made using an induction furnace. The crystals were annealed at 1550 K for one week, followed by annealing at 873 K for another week. The powder X-ray diffraction measurements confirmed that the Co$_2$MnAl single crystals grown using the above method possesses a cubic structure. Since Co$_2$MnAl has multiple structural phases with different types of disorders as mentioned in the main text and only the L2$_1$ phase is predicted to host a FM WSM phase, we performed detailed transmission electron microscopy (TEM) analyses to identify its structure phase. The TEM sample was prepared using a Thermo Fisher Helios NanoLab Dual-Beam focused ion beam system. The HAADF-STEM images are taken with the Thermo Fisher Titan S/TEM equipped with a spherical aberration corrector, and it was operating at 300 kV accelerating voltage with a probe convergence angle of 30 mrad. Magnetization properties were measured using a superconducting quantum interference device (SQUID VSM, Quantum Design). The electrical transport properties were measured using a standard four probe method in a Physical property measurement system (PPMS, Quantum Design).

Here are references for Fig.~\ref{Fig:TableS1}.
\begin{enumerate}
\item Suzuki, T. et al. Large anomalous Hall effect in a half-Heusler antiferromagnet. Nat. Phys.12, 1119-1123 (2016).
\item  Ye, L. et al. Massive Dirac fermions in a ferromagnetic kagome metal. Nature 555, 638 (2018).
\item Manyala, N. et al. Large anomalous Hall effect in a silicon-based magnetic semiconductor. Nat. Mater. 3, 255 (2004).
\item Imort, I. M. et al. Anomalous Hall effect in the Co-based Heusler compounds Co2FeSi and Co2FeAl. J. Appl. Phys. 111, 07D313 (2012).
\item Liu, E. et al. Giant anomalous Hall effect in a ferromagnetic kagome-lattice semimetal.  Nat. Phys. 14, 1125–1131 (2018).
\item Nakatsuji, S. et al. Large anomalous Hall effect in a non-collinear antiferromagnet at room temperature. Nature 527, 212 (2015). 
\item Kim, K. et al. Large anomalous Hall current induced by topological nodal lines in a  ferromagnetic van der Waals semimetal. Nat. Mater. 17, 794-799 (2018).
\item Yu, J. et al. Magnetotransport and magnetic properties of molecular-beam epitaxy L10 FePt thin films. J. Appl. Phys. 87, 6854-6856 (2000). 
\item Zeng, C. et al. Linear magnetization dependence of the intrinsic anomalous Hall effect. Phys. Rev. Lett. 96, 037204 (2006).
\item Wu, F. et al. Electrical transport properties of perpendicular magnetized Mn-Ga epitaxial films. Appl. Phys. Lett. 96, 042505 (2010).
\item Miyasato, T. et al. Crossover behavior of the anomalous Hall effect and anomalous Nernst effect in itinerant ferromagnets. Phys. Rev. Lett. 99, 086602 (2007).
\item Dijkstra, J. et al. Band-structure calculations of Fe1/3TaS2 and Mn1/3TaS2 , and transport and magnetic properties of Fe0.28TaS2. J. Phys.: Condens. Matter 1, 6363 (1989).
\item Fang, Z. et al. The anomalous Hall effect and magnetic monopoles in momentum space. Science 302, 92-95 (2003).
\item Thiyagarajah, N. et al. Giant spontaneous Hall effect in zero-moment Mn2RuxGa. Appl. Phys. Lett. 106, 122402 (2015).
\item Kim, T. et al. Spontaneous Hall effect in amorphous Tb–Fe and Sm–Fe thin films. J. Appl. Phys. 89, 7212-7214 (2001).
\item Nayak, A. K. et al. Large anomalous Hall effect driven by a nonvanishing Berry curvature in the noncolinear antiferromagnet Mn3Ge. Sci. Adv. 2, e1501870 (2016).
\item  Kim, T. W. and Gambino, R. J. Composition dependence of the Hall effect in amorphous TbxCo1-x thin films. J. Appl. Phys. 87, 1869-1873 (2000).
\item Pu, Y. et al. Mott relation for anomalous Hall and Nernst effects in Ga1-xMnxAs ferromagnetic semiconductors. Phys. Rev. Lett. 101, 117208 (2008).
\item Sakai, A., et al. Giant anomalous Nernst effect and quantum-critical scaling in a ferromagnetic semimetal. Nat. Phys. 14, 1119-1124 (2018).

\end{enumerate}

\end{document}